\documentclass[twocolumn,showkeys,aps,prb,showpacs]{revtex4-1}
\usepackage{graphicx}
\usepackage[CJKbookmarks,dvipdfm,colorlinks,linkcolor=blue,citecolor=blue]{hyperref}

\begin{document}

\title{Large piezoelectric coefficients combined with high electron  mobilities in Janus monolayer XTeI (X=Sb and Bi): a first-principle study}

\author{San-Dong Guo,  Xiao-Shu Guo, Zhao-Yang Liu and Ying-Ni Quan}
\affiliation{School of Electronic Engineering, Xi'an University of Posts and Telecommunications, Xi'an 710121, China}
\begin{abstract}
The absence of both the inversion symmetry and out-of-plane mirror symmetry together with spin-orbit coupling (SOC) can induce novel electronic and piezoelectric properties.  In this work, the  piezoelectric properties along with carrier mobilities of Janus monolayer XTeI (X=Sb and Bi) are studied by
density functional theory (DFT).  By using generalized gradient approximation (GGA) plus SOC, they are found to be indirect gap semiconductors with  the Rashba spin splitting. The piezoelectric tensors  of Janus monolayer XTeI (X=Sb and Bi) are reported by using density functional perturbation theory (DFPT). Due to lacking both the inversion symmetry and out-of-plane mirror symmetry for Janus monolayer XTeI (X=Sb and Bi),
both in-plane and out-of-plane piezoelectric effects can be observed, and the large  piezoelectric coefficients are predicted (e.g. $d_{11}$=12.95 pm/V for SbTeI and  $d_{11}$=8.20 pm/V for BiTeI), which are
comparable and even higher than ones of many other two-dimensional (2D) materials and other well-known bulk piezoelectric materials, especially for out-of-plane piezoelectric coefficients. With GGA+SOC, the high electron carrier  mobilities are obtained, and the electron mobility of BiTeI along  armchair direction  reaches  up to about 1319 $\mathrm{cm^2V^{-1}s^{-1}}$.  The carrier mobility shows a rather pronounced anisotropy between electron and hole/armchair and zigzag directions. It is found that tensile strain can improve the  piezoelectric coefficients $d_{11}$ of Janus monolayer XTeI (X=Sb and Bi).
For example, at 4\%  strain, the $d_{11}$ of SbTeI (BiTeI)  is up to 20.12 pm/V (11.48 pm/V), compared with unstrained 12.95 pm/V (8.20 pm/V).
 Our works imply Janus monolayer XTeI (X=Sb and Bi)
have potential  applications in  flexible electronics and piezoelectric devices, and  can stimulate further experimental works.

\end{abstract}

\pacs{71.20.-b, 77.65.-j, 72.15.Jf, 78.67.-n ~~~~~~~~~~~~~~~~~~~~~~~~~~~~~~~~~~~Email:sandongyuwang@163.com}

\maketitle

\section{Introduction}
Compared to their bulk counterparts, 2D materials can show some
unique or surprising properties, like high mechanical strength\cite{q1}, high photocatalytic activity\cite{q2}, remarkable
optical and magnetic properties\cite{q3}.
Due to potential nanoscale piezoelectric
applications, the piezoelectricity in 2D materias has
attracted growing interest\cite{q4}, which can be used in sensors, actuators and
energy sources.  A piezoelectric material must be non-centrosymmetric,  and should have a band gap for
prohibiting current leakage. For 2D materials, the reduction in dimensionality  leads to that their inversion symmetry
is often eliminated,  allowing them to become piezoelectric.
Due to the inversion asymmetry in the
2H phase,  the monolayer $\mathrm{MoS_2}$ as a typical example have been  proved to possess piezoelectricity experimentally ($e_{11}$=2.9$\times$$10^{-10}$ C/m)\cite{q5,q6}.
A large in-plane piezoelectricity and
additional out-of-plane piezoelectricity have been predicted in 2D Janus transition metal dichalchogenides (TMD)\cite{q7}, which can directly break the inversion symmetry and
out-of-plane mirror symmetry. Recently, the existence of vertical dipoles has been experimentally proved in the Janus MoSSe monolayer,
which shows  an intrinsic vertical piezoelectric response\cite{q8}.
In theory, the piezoelectric properties of  lots of 2D materials  have been reported\cite{q7,q9,q10,q11,q12}, like  metal dichalcogenides, group IIA and IIB metal oxides, group III-V semiconductors and Janus TMD.
\begin{figure}
  \includegraphics[width=7.0cm]{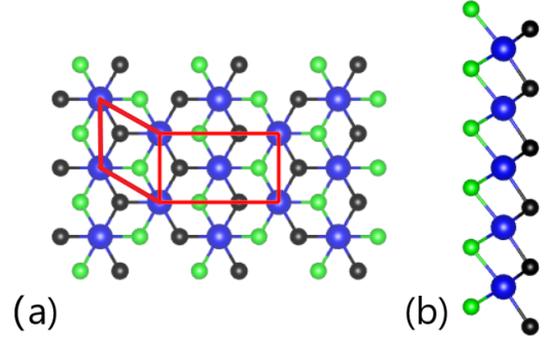}
  \caption{(Color online) The  crystal structure of Janus monolayer XTeI (X=Sb and Bi): top view (a) and side view (b). The large blue balls represent X atoms, and the  small black/green balls for Te/I atoms. The  rhombus primitive cell  and the rectangle supercell are shown.}\label{t0}
\end{figure}

\begin{figure*}
  \includegraphics[width=12cm]{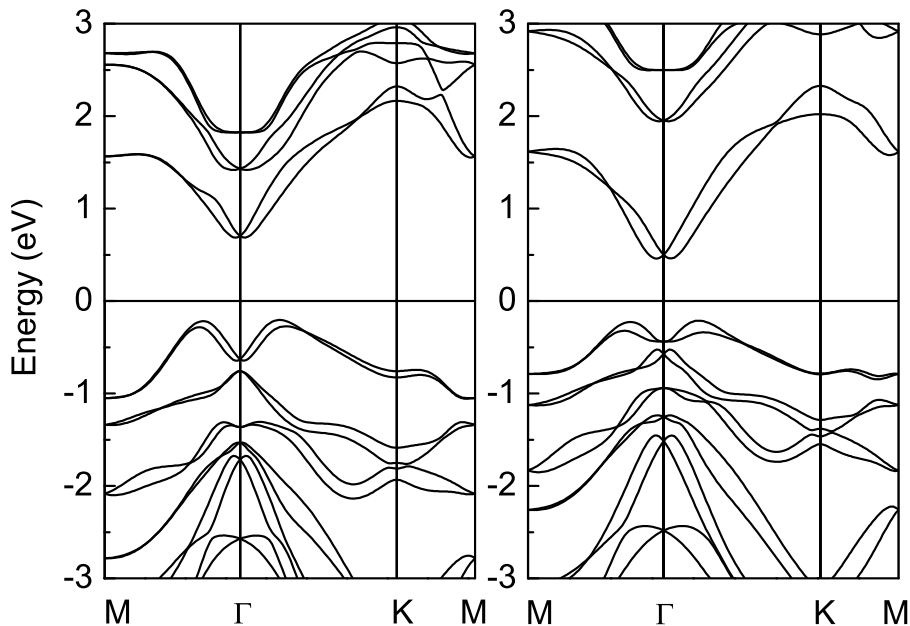}
\caption{The GGA+SOC energy band structures  of  SbTeI (Left) and BiTeI (Right) monolayers. }\label{t1-1}
\end{figure*}

Janus monolayer SbTeI and BiTeI have been predicted theoretically with  a low formation energy,  and they are dynamically
stable\cite{q13,q14}.  The bulk SbTeI or BiTeI  has a layered structure along its crystallographic c axis, and a Bi or Sb atom is
sandwiched between one Te and one I, forming a triple layer. Within each triple layer, these atoms have strong
chemical bonding, but the adjacent triple layers are weakly coupled via the van der Waals interactions. Thus, Janus monolayer SbTeI and BiTeI
can be exfoliated  from their bulks. The Rashba
spin splitting in  monolayer SbTeI or BiTeI has been reported  with a sizable Rashba coupling parameter of
1.39 eV$\mathrm{{\AA}}$ or 1.97 eV$\mathrm{{\AA}}$\cite{q14}, and they are potentially useful for spintronics, like spin-polarized field effect transistor. Recently, we have predicted that Janus monolayer SbTeI and BiTeI may be potential 2D thermoelectric materials due to very low lattice
thermal conductivities\cite{q15}. However, to the best of our knowledge, no studies have been reported on piezoelectric properties of Janus monolayer SbTeI and BiTeI. In this work, inspiring from their special Janus structure of monolayer SbTeI or BiTeI, their  piezoelectric properties along with carrier mobilities  are studied by density functional theory (DFT). Calculated results show  they have the large  piezoelectric coefficients (e.g. $d_{11}$=12.95 pm/V for SbTeI and  $d_{11}$=8.20 pm/V for BiTeI), which are
comparable and even higher than ones of many other 2D materials\cite{q7,q9,q11,q12} and other well-known bulk piezoelectric materials\cite{aln,aln-1,aln-2,aln-3,aln-4}, especially for out-of-plane piezoelectric coefficients with respect to other ones of 2D materials. The high electron carrier  mobilities are also predicted,  for example about 1319 $\mathrm{cm^2V^{-1}s^{-1}}$  for the electron mobility of BiTeI along  armchair direction. It is proved that the tensile strain can make the $d_{11}$ of SbTeI (BiTeI)  to be up to 20.12 pm/V (11.48 pm/V) from unstrained 12.95 pm/V (8.20  pm/V), increased by 55\% (40\%). These results make us believe that Janus monolayer SbTeI and BiTeI may be potential 2D piezoelectric materials,  and can
stimulate further experimental  synthesis of these monolayers.

\begin{table}
\centering \caption{For SbTeI (BiTeI) monolayer, the lattice constants $a_0$ ($\mathrm{{\AA}}$), the elastic constants $C_{ij}$ ($\mathrm{Nm^{-1}}$), shear modulus
$G_{2D}$ ($\mathrm{Nm^{-1}}$),  Young's modulus $C_{2D}$  ($\mathrm{Nm^{-1}}$),  Poisson's ratio $\nu$,  the GGA+SOC gaps  (eV) and the Rashba energy $E_{R}$ (meV). }\label{tab0}
  \begin{tabular*}{0.48\textwidth}{@{\extracolsep{\fill}}cccc}
  \hline\hline
$a_0$& $C_{11}=C_{22}$ &  $C_{12}$& $G_{2D}$\\\hline
4.32 (4.42) &28.52 (26.91)&7.72 (7.93)&10.4 (9.49)\\\hline\hline
$C_{2D}$& $\nu$& $Gap$& $E_{R}$\\\hline
26.43 (24.57)&0.271 (0.295)&0.89 (0.67) & 17.2 (40.1)\\\hline\hline
\end{tabular*}
\end{table}

\section{Computational detail}
 Based on DFT\cite{1}, our calculations are performed using the projected augmented wave
(PAW) method with a kinetic cutoff energy of 500 eV, as implemented in
the  VASP package\cite{pv1,pv2,pv3}.  The popular GGA of Perdew, Burke and  Ernzerhof  (GGA-PBE)\cite{pbe} is used as the exchange-correlation potential, and the SOC was included self-consistently. For the Janus  monolayer SbTeI and BiTeI, a vacuum spacing of
 more than 18 $\mathrm{{\AA}}$ along the z direction is included to avoid interactions
between two neighboring images. The total energy  convergence criterion is set
to $10^{-8}$ eV. The geometry optimization was considered
to be converged with the residual force on each atom being less than 0.001 $\mathrm{eV.{\AA}^{-1}}$.
To obtain the piezoelectric properties of the Janus  monolayer SbTeI and BiTeI, the elastic stiffness tensor $C_{ij}$ are calculated by using the finite difference method (FDM), and the piezoelectric stress coefficients $e_{ij}$ are calculated by  DFPT method\cite{pv6} by  the VASP package.
Both elastic and piezoelectric stress  coefficients,  including both the ionic
and electronic components, have been calculated. Within FDM and DFPT, the
 the electronic and ionic contribution to
the elastic and  piezoelectric stress coefficient can be given  directly from VASP code.
The 2D elastic coefficients $C^{2D}_{ij}$
 and   piezoelectric stress coefficients $e^{2D}_{ij}$
have been renormalized by the the length of unit cell along z direction ($Lz$):  $C^{2D}_{ij}$=$Lz$$C^{3D}_{ij}$ and $e^{2D}_{ij}$=$Lz$$e^{3D}_{ij}$.

\begin{figure*}
  \includegraphics[width=12cm]{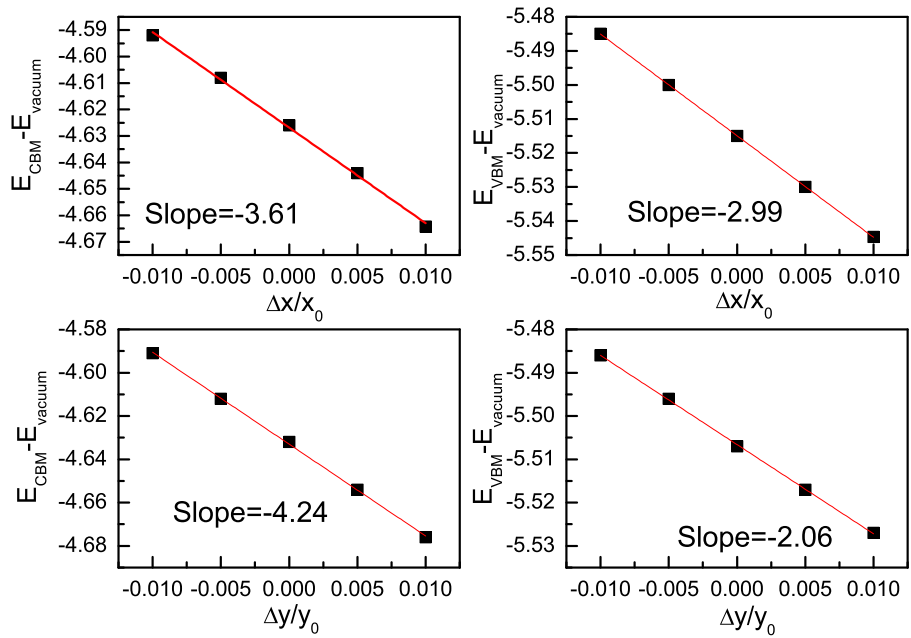}
  \caption{(Color online)With respect to the vacuum energy, the GGA+SOC band energies of the VBM and CBM of SbTeI monolayer  as a function of lattice dilation along both x and y directions. The red solid lines are linear fitting curves, and the fitted slopes are given,  corresponding to the DP.}\label{sb}
\end{figure*}

\section{Electronic structures}
The geometric structure of the Janus  XTeI (X=Sb and Bi) is plotted in \autoref{t0},  which consists of of three atomic layers with X sandwiched between the Te and I layers. The structure  is similar to  Janus TMD monolayer PtSSe\cite{q7-0} and SnSSe\cite{q7-2}.
Both rhombus primitive cell  and rectangle supercell are shown, and calculating  carrier mobilities and piezoelectric coefficients need to use rectangle supercell.  The armchair and zigzag directions of the rectangular supercell are defined as x and y directions, respectively.
Their symmetry is  $P3m_1$,  which lacks the reflection symmetry with respect to the central  X atoms.
The lattice constant of the Janus  monolayer SbTeI and BiTeI are 4.32  $\mathrm{{\AA}}$ and  4.42 $\mathrm{{\AA}}$, which agree well previous theoretical values\cite{q14,q15}. It have been proved that the SOC has very important effects on their electronic structures\cite{q13,q14,q15}, thus  only GGA+SOC energy bands of monolayer SbTeI and BiTeI are plotted in \autoref{t1-1}.
Around the Fermi level,  all the
bands are dominated by the p-orbits with  conduction band
minimum (CBM) being  Sb/Bi-p and valence band maximum (VBM) being Te-p\cite{q13}.
The Janus  monolayer SbTeI  (BiTeI) is found to be a semiconductors
with an indirect band gap of 0.89 eV (0.67 eV) with GGA+SOC.  The  CBM is located at slightly away from
the $\Gamma$ point along the $\Gamma$-K or $\Gamma$-M paths, and the VBM is  between the $\Gamma$ and K or M points.
The Rashba-like and Zeeman-like band splitting can be observed, which is due to the absence of vertical mirror symmetry
combined with the SOC effect. The Rashba spin
splitting in the  lowest conduction band at the high-symmetric  $\Gamma$ point can be clearly seen, and their Rashba energy, defined
as  the band crossing point of the conduction bands at the $\Gamma$ point  minus the energy of  the CBM,  are
17.2 meV and 40.1 meV, respectively, which are consistent
with the previously reported values\cite{q13,q14,q15}.
\begin{table}
\centering \caption{Piezoelectric coefficients $e_{11}(d_{11})$ and $e_{31} (d_{31})$ of SbTeI and BiTeI monolayers, along with ones of some typical 2D
 materials, like $\mathrm{MoS_2}$\cite{q9}, $\mathrm{WS_2}$\cite{q9}, ZnO\cite{q9}, MoSSe\cite{q7}$^($\cite{q7-2}$^)$ and MoSTe\cite{q7}$^($\cite{q7-1}$^)$. The unit is $10^{-10}$C/m  for $e_{ij}$ (pm/V for $d_{ij}$). }\label{tab-y}
  \begin{tabular*}{0.48\textwidth}{@{\extracolsep{\fill}}ccccc}
  \hline\hline
Name & $e_{11}$ & $d_{11}$& $e_{31}$&$d_{31}$\\\hline\hline
SbTeI&2.69& 12.95&-0.13&-0.37\\\hline
BiTeI& 1.56&     8.20   &-0.23 & -0.66                                \\\hline\hline
$\mathrm{MoS_2}$&3.62& 3.65&&\\\hline
$\mathrm{WS_2}$&2.43& 2.12&&\\\hline
$\mathrm{ZnO}$&2.66& 8.65&&\\\hline
$\mathrm{MoSSe}$&3.74 (3.89)& 3.76 (4.24)&0.032 (0.42) & 0.02 (0.29)\\\hline
$\mathrm{MoSTe}$&4.53 (4.5)& 5.04 (5.1)& 0.038 (0.5) &  0.028 (0.4) \\\hline\hline
\end{tabular*}
\end{table}

\begin{table}
\centering \caption{The ionic contribution and electronic contribution to $e_{ij}$ of Janus monolayer SbTeI and BiTeI. }\label{tab-cc}
  \begin{tabular*}{0.48\textwidth}{@{\extracolsep{\fill}}cccccc}
  \hline\hline
Name & $e_{11}$ & $e_{31}$& &$e_{11}$&$e_{31}$\\\hline\hline
     & Ionic contribution & & &Electronic contribution &\\\hline
SbTeI&2.23& -0.17&&0.46&0.04\\\hline
BiTeI& 1.26&     -0.17   &&0.30 & -0.06                               \\\hline\hline
\end{tabular*}
\end{table}

\section{Piezoelectric properties}
For both in-plane and out-of plane
directions, SbTeI and BiTeI monolayers clearly lack an inversion center, which means
both the in-plane and out-of-plane piezoelectricity can be produeced, when  a tension along x
axis is applied. The third-rank piezoelectric stress tensors  $e_{ijk}$ and strain tensor $d_{ijk}$ from the sum of ionic
and electronic contributions are defined as:
 \begin{equation}\label{pe0}
      e_{ijk}=\frac{\partial P_i}{\partial \varepsilon_{jk}}=e_{ijk}^{elc}+e_{ijk}^{ion}
 \end{equation}
and
 \begin{equation}\label{pe0-1}
   d_{ijk}=\frac{\partial P_i}{\partial \sigma_{jk}}=d_{ijk}^{elc}+d_{ijk}^{ion}
 \end{equation}
In which $P_i$, $\varepsilon_{jk}$ and $\sigma_{jk}$ are polarization vector, strain and stress,respectively.
For 2D materials, the  plane stress and plane
strain conditions are assumed($\varepsilon_{jk}$=$\sigma_{ij}$=0 for i=3 or j=3)\cite{q9,q10,q11,q12}.
 Due to a $3m$ point-group symmetry of SbTeI and BiTeI monolayers,  by employing Voigt notation,
 the  piezoelectric stress   and strain tensors become:
 \begin{equation}\label{pe1-1}
 e=\left(
    \begin{array}{ccc}
      e_{11} & -e_{11} & 0 \\
     0 & 0 & -e_{11} \\
      e_{31} & e_{31} & 0 \\
    \end{array}
  \right)
    \end{equation}

  \begin{equation}\label{pe1-2}
  d= \left(
    \begin{array}{ccc}
      d_{11} & -d_{11} & 0 \\
      0 & 0 & -2d_{11} \\
      d_{31} & d_{31} &0 \\
    \end{array}
  \right)
\end{equation}
where $e/d_{11}$ and $e/d_{31}$ represent the in-plane and out-plane piezoelectric
stress/strain components, respectively. The $e_{11}$ is induced  by
uniaxial in-plane uniaxial strain, while $e_{31}$ is produced by uniaxial strain along the
a-axis. The
$e_{31}$ is non-zero, which is due to the absence
of an inversion centre between the Te and I layers. The $e_{ij}$ can be calculated by DFPT, and derive the values of
$d_{ij}$ using the relation:
 \begin{equation}\label{pe1-3}
    e=dC
 \end{equation}
 and the  elastic tensor $C$ can be expressed as:
\begin{equation}\label{pe1-4}
   C=\left(
    \begin{array}{ccc}
      C_{11} & C_{12} & 0 \\
     C_{12} & C_{11} &0 \\
      0 & 0 & (C_{11}-C_{12})/2 \\
    \end{array}
  \right)
\end{equation}
The $C_{ij}$ can be attained by FDM.
Here, the  $d_{11}$ and $d_{31}$ are derived by  \autoref{pe1-1}, \autoref{pe1-2}, \autoref{pe1-3} and \autoref{pe1-4}:
\begin{equation}\label{pe2-7}
    d_{11}=\frac{e_{11}}{C_{11}-C_{12}}~~~and~~~d_{31}=\frac{e_{31}}{C_{11}+C_{12}}
\end{equation}

In general, a piezoelectric material should be a semiconductor for
prohibiting current leakage, which has been proved from energy bands structures from \autoref{t1-1}.
The elastic stiffness coefficients of $C_{11}$and $C_{12}$ of SbTeI and BiTeI monolayers are firstly calculated, and
 the shear modulus $G_{2D}$,  Young's modulus $C_{2D}$ and  Poisson's ratio $\nu$ are attained from calculated $C_{ij}$\cite{ela},
which are listed in \autoref{tab0}. These elastic  coefficients $C_{ij}$ satisfy the Born criteria
of 2D hexagonal structure\cite{ela} ($C_{11}$>0 and $C_{11}$-$C_{12}$)>0, proving  their mechanical stabilities.
 These values  are obviously smaller than those of
other 2D materials,  like  monolayer TMDs\cite{q7,q9}, which  means that SbTeI and BiTeI  monolayers are more flexible than other 2D materials. Therefore, they  are very favorable for novel flexible piezotronics.

For $e_{ij}$, the nonprimitive orthorhombic unit cell  is used  as the computational unit cell.  The calculated in-plane $e_{11}$ ($d_{11}$) of SbTeI monolayer are 2.69$\times$$10^{-10}$ C/m (12.95 pm/V), and   1.56$\times$$10^{-10}$ C/m (8.20 pm/V) for BiTeI monolayer.
Their  $d_{11}$ is much higher than that of the widely used bulk materials\cite{aln,aln-1,aln-2,aln-3,aln-4}
like $\alpha$-quartz ($d_{11}$=2.3 pm/V),  wurtzite-AlN ($d_{33}$=5.1 pm/V) and wurtzite-GaN ($d_{33}$=3.1 pm/V), and are found to be
comparable to or even superior than other 2D materials\cite{q7,q9,q11,q12,q7-2,q7-1}, such as $\mathrm{MoS_2}$, $\mathrm{WS_2}$, ZnO, MoSSe and MoSTe.
The related data are listed \autoref{tab-y}.
These  imply  that the SbTeI and BiTeI  monolayers
may have large in-plane piezoelectric effects  with a uniaxial strain being applied.

For 2D materials, a large out-of-plane piezoelectric response is
highly desired, since it is compatible with the nowadays bottom/top gate technologies.
The calculated  out-of-plane piezoelectric strain coefficient ($d_{31}$) has large value of  -0.37 pm/V for SbTeI  monolayer,
 and -0.66 pm/V for  BiTeI monolayer.
The out-of-plane piezoelectricity $d_{31}$ of BiTeI monolayer is obviously
higher than ones of other 2D  materials, like
functionalized h-BN (0.13 pm/V)\cite{y1}, MoSSe (0.29 pm/V)\cite{q7-2}, MoSTe (0.4 pm/V)\cite{q7-1}, Janus group-III materials (0.46 pm/V)\cite{y2} and $\alpha$-$\mathrm{In_2Se_3}$ (0.415 pm/V)\cite{y3}.  It is noted that the $e_{31}$ and  $d_{31}$ of MoSSe and MoSTe are very different between ref.\cite{q7} and ref.\cite{q7-2,q7-1}, which are likely due to different computational method. In ref.\cite{q7-2,q7-1}, the $e_{ij}$ are calculated by DFPT, while these are attained by evaluating the change of unit-cell polarization after imposing small strain in ref.\cite{q7}. It is clearly seen that the $e_{11}$ are very close for different method, but the $e_{31}$ by polarization-strain relation is  smaller by 1 order  of magnitude compared to one by DFPT.
The large out-of-plane piezoelectric effect  makes SbTeI and BiTeI monolayers have potential
applications on ultrathin piezoelectric  devices.
\begin{table*}
\centering \caption{The calculated  elastic modulus ($C_{2D}$), effective mass ($m^*$) using GGA+SOC, deformation potential ($E_l$), carrier mobility ($\mu_{2D}$) [300 K]   and relaxation time ($\tau$) [300 K] of SbTeI and BiTeI monolayers.}\label{tab-u2d}
  \begin{tabular*}{0.96\textwidth}{@{\extracolsep{\fill}}cccccccc}
  \hline\hline
&Carrier type&    &$C_{2D}$ ($\mathrm{Nm^{-1}}$) & $m^*$ & $E_l$ (eV)& $\mu_{2D}$ ($\mathrm{cm^2V^{-1}s^{-1}}$)&$\tau$ (s)\\\hline\hline
&Electrons   & x&  26.43&       0.21&       -3.61          & 739.74&  $8.88\times10^{-14}$     \\
SbTeI&            &y&    26.43&       0.36 &      -4.24         & 311.70   & $6.43\times10^{-14}$                                         \\
&Holes   & x&       26.43&      -1.86&     -2.99          &34.97     & $3.70\times10^{-14}$                                           \\
&             &y&   26.43&      -0.50&     -2.06           &272.59      & $7.80\times10^{-14}$                             \\\hline\hline
&Electrons   & x&  24.57&        0.19&       -2.62        & 1319.24&  $1.40\times10^{-13}$   \\
BiTeI&            &y&    24.57&       0.52 &      -3.04          & 351.18   & $1.04\times10^{-13}$                                            \\
&Holes   & x&       24.57&      -1.36&     -2.06           &102.92     & $7.95\times10^{-14}$                                           \\
&             &y&   24.57&      -0.57&     -1.66          &376.58      & $1.22\times10^{-13}$                             \\\hline\hline
\end{tabular*}
\end{table*}

It is found that  the magnitude of the ionic contribution is
larger than that of the electronic contribution. For monolayer SbTeI (BiTeI),  the ionic contribution accounts for 83\% (81\%) of $e_{11}$, and
100\% [The contribution between ionic and electronic one is opposite.] (75\%) of $e_{31}$. The ionic contribution and electronic contribution to $e_{ij}$ are listed in \autoref{tab-cc}. Thus, it is very important for calculating the piezoelectric stress tensors to consider relaxed-ion, not clamped-ion (only electronic contribution). In fact, the real piezoelectric stress tensor $e_{ij}$ (a relaxed-ion
result) is the sum of ionic  and electronic contributions.

\section{Carrier mobility}
To  evaluate the performance of monolayer SbTeI and BiTeI for   great potential in  piezoelectric and electronic application,
the carrier mobilities of their electron/hole transport  can be calculated by the deformation potential (DP) theory  proposed by Bardeen and Shockley\cite{dp}. In this theory, the expression of the carrier
mobility of a 2D material  ($\mu_{2D}$)  is given:
\begin{equation}\label{u2d}
  \mu_{2D}=\frac{e\hbar^3C_{2D}}{K_BTm^*m_dE_l^2}
\end{equation}
where $e$, $\hbar$ and $K_B$ are the electron charge, the reduced Planck
constant and  the Boltzmann constant; the  $T$ and   $m^*$ are temperature and  effective mass in the transport direction, and $m_d=\sqrt{m_xm_y}$ is the average effective mass.
The $C_{2D}$ is the elastic modulus  derived from elastic constants $C_{ij}$, and the relationship between them is\cite{ela}:
\begin{equation}\label{e1}
C_{2D}=\frac{C_{11}^2-C_{12}^2}{C_{11}}
\end{equation}
In addition, $E_l$ represents the DP constant
of the VBM for a hole or the CBM for an electron along the transport direction, which is defined by $E_l=\Delta E/\delta$
, where $\Delta E$ is the energy shift of the band edge of CBM or VBM with respect
to the vacuum level.  These data are calculated with strain from -0.01 to 0.01, and the step $\delta$=0.005.
When the $\mu_{2D}$ is attained,  the  relaxation
time $\tau$ of  a hole at VBM or  an electron at CBM along the transport direction can be calculated by:
\begin{equation}\label{t}
    \tau=\mu_{2D}m^*/e
\end{equation}

\begin{figure*}
  \includegraphics[width=12cm]{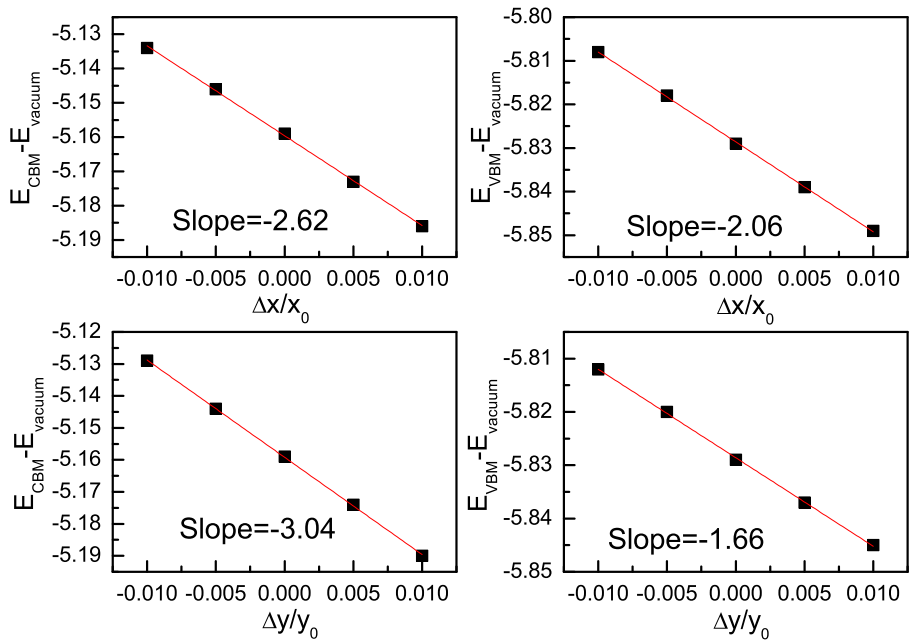}
  \caption{(Color online)With respect to the vacuum energy, the GGA+SOC band energies of the VBM and CBM of BiTeI monolayer  as a function of lattice dilation along both x and y directions. The red solid lines are linear fitting curves, and the fitted slopes are given,  corresponding to the DP.}\label{bi}
\end{figure*}

The rectangular supercell  with armchair and zigzag being defined as x and y directions in \autoref{t0} is used to study the  carrier mobilities of SbTeI and BiTeI monolayers, and the  300 K (room temperature) is
used in the mobility calculations. It is noteworthy that calculating effective masses and the DP constant  need consider SOC. Firstly, the effective masses of CBM and VBM along armchair and zigzag directions are calculated by:
\begin{equation}
(m_{l})^{-1}=\frac{1}{\hbar^2}\frac{\partial^2E(k)}{\partial k_l^2}
\end{equation}
In which  $E(k)$ is the dispersion of the lowest conduction band or the highest valence band.
For both monolayer SbTeI and BiTeI, the calculated effective masses of electrons are smaller than ones of holes, and the effective masses along armchair direction are smaller than ones along zigzag direction for electrons, but that is opposite for holes.
The band energies of the VBM and CBM  with respect to the vacuum energy as a function of $\Delta x/x$ and $\Delta y/y$ are plotted in \autoref{sb} for SbTeI monolayer and in \autoref{bi} for BiTeI monolayer. The DP constant $E_l$ can be attained by linear fitting, and the related slopes are  DP constant $E_l$. Based on \autoref{u2d}, the carrier mobilities of SbTeI and BiTeI monolayers, including the electrons and holes  along x and y directons,  are calculated, and they are shown \autoref{tab-u2d}.
The electron mobilities of monolayer SbTeI and BiTeI are in the range of 311.70-1319.24
$\mathrm{cm^2V^{-1}s^{-1}}$, which are larger than those of
holes (34.97-376.58 $\mathrm{cm^2V^{-1}s^{-1}}$). This mainly is because the effective masses of electron are smaller than ones of holes.
Furthermore, the
carrier mobilities  between  electrons and holes/between armchair and zigzag directions show very strong anisotropy.
According to \autoref{t}, the $\tau$ can be calculated from effective masses and  carrier mobilities, which  is useful for attaining the electronic  transport coefficients to study their electronic  transport  properties.
\begin{table}
\centering \caption{For Janus monolayer  SbTeI and BiTeI,  the elastic constants $C_{ij}$ ($\mathrm{Nm^{-1}}$), shear modulus
$G_{2D}$ ($\mathrm{Nm^{-1}}$),  Young's modulus $C_{2D}$  ($\mathrm{Nm^{-1}}$), along with ones of some typical 2D
 materials, like $\mathrm{MoS_2}$\cite{q9}, $\mathrm{WS_2}$\cite{q9}, ZnO\cite{q9}, MoSSe\cite{q7} and MoSTe\cite{q7}.}\label{tab-c}
  \begin{tabular*}{0.48\textwidth}{@{\extracolsep{\fill}}ccccc}
  \hline\hline
Name & $C_{11}$ & $C_{12}$& $G_{2D}$&$C_{2D}$\\\hline\hline
SbTeI&28.5& 7.7       &10.4   & 26.4 \\\hline
BiTeI& 26.9&   7.9   & 9.5   &   24.6                              \\\hline\hline
$\mathrm{MoS_2}$&130.3& 31.0&  49.7&   122.9\\\hline
$\mathrm{WS_2}$&146.0& 31.6& 57.2&139.2\\\hline
$\mathrm{ZnO}$&92.6& 61.9&15.35&51.2\\\hline
$\mathrm{MoSSe}$&126.8& 27.4&49.7 &120.9\\\hline
$\mathrm{MoSTe}$&112.7&22.7&45 & 108.1\\\hline\hline

\end{tabular*}
\end{table}

\begin{figure*}
  \includegraphics[width=13cm]{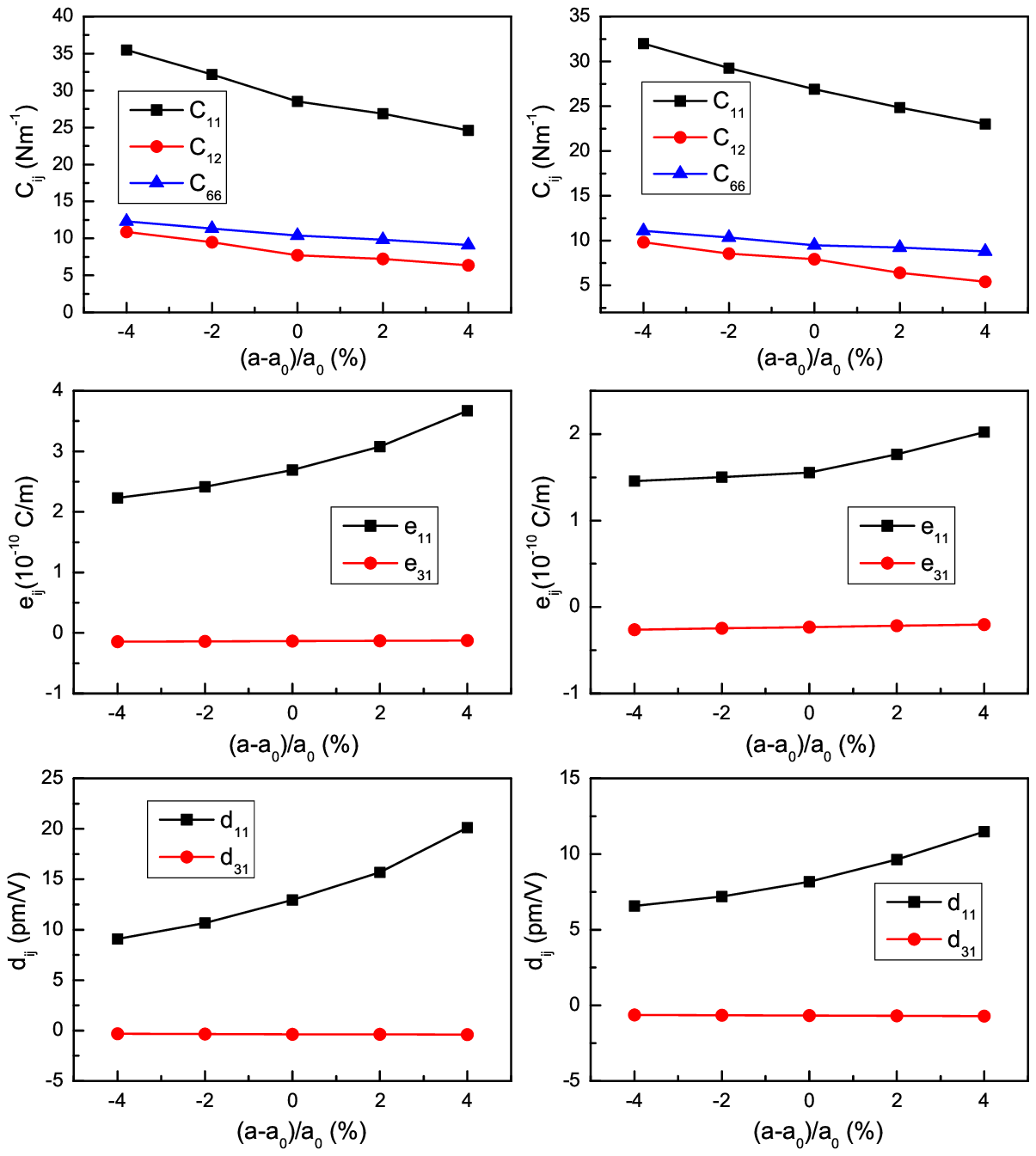}
\caption{(Color online) For monolayer SbTeI (Left) and BiTeI (Right), the  elastic constants $C_{ij}$, piezoelectric coefficients $e_{ij}$  and  $d_{ij}$    with the application of  biaxial strain (-4\%-4\%). }\label{s}
\end{figure*}

\section{Discussions and Conclusion}
From \autoref{tab-y}, the $e_{11}$ of Janus monolayer SbTeI and BiTeI are found to be
comparable to or even smaller than ones of  $\mathrm{MoS_2}$, $\mathrm{WS_2}$, ZnO, MoSSe and MoSTe\cite{q7,q9,q7-2,q7-1}, but their
$d_{11}$ are larger than  ones of those 2D materials, except for ZnO. To find out underlying reasons,  the elastic constants $C_{ij}$, shear modulus
$G_{2D}$,  Young's modulus $C_{2D}$ of Janus monolayer SbTeI and BiTeI  along with ones of  $\mathrm{MoS_2}$\cite{q9}, $\mathrm{WS_2}$\cite{q9}, ZnO\cite{q9}, MoSSe\cite{q7} and WSTe\cite{q7} are listed in \autoref{tab-c}. Based on \autoref{pe2-7}, the $d_{11}$ is inversely proportional to $G_{2D}$ [$G_{2D}$=($C_{11}-C_{12}$)/2]. It is clearly seen that $G_{2D}$ of Janus monolayer SbTeI and BiTeI are very  smaller than ones of other 2D materials listed in \autoref{tab-c}, which leads to larger $d_{11}$. It is also noted that $G_{2D}$ of monolayer ZnO is relatively small, along with
relatively large $e_{11}$, which gives rise to large $d_{11}$ (8.65 pm/V).  The $d_{31}$ is inversely proportional to ($C_{11}+C_{12}$), and both $C_{11}$ and $C_{12}$ of Janus monolayer SbTeI and BiTeI  are smaller than other ones from \autoref{tab-c}. Thus, the $e_{31}$ of Janus monolayer SbTeI and BiTeI are found to be
 smaller than ones of  MoSSe and MoSTe\cite{q7-2,q7-1} using the same DFPT method, but their
$d_{31}$ are comparable to or even larger than  ones of MoSSe and MoSTe.

Strain is a very effective  measures to tune the electronic,  topological and  transport properties of 2D materials.
Here, we also investigate  the small  biaxial strain (-4\% to 4\%) effects on  piezoelectric properties of monolayer SbTeI and BiTeI.
The  elastic constants $C_{ij}$, piezoelectric coefficients $e_{ij}$  and  $d_{ij}$ of   monolayer SbTeI and BiTeI  as a function of strain
are plotted in \autoref{s}. When the  strain changes from -4\% to 4\%,
all  $C_{ij}$ of both monolayer SbTeI and BiTeI show a  monotone decrease, which is in favour of improving  piezoelectric effects according to \autoref{pe2-7}.
It is found that  piezoelectric coefficient $e_{11}$  increases with strain from -4\% to 4\%, which leads to enhanced $d_{11}$ along with reduced $C_{66}$. At
4\%  strain, the $d_{11}$ of SbTeI (BiTeI)  reaches up to 20.12 pm/V (11.48 pm/V) from unstrained 12.95 pm/V (8.20  pm/V), increased by 55\% (40\%).
Calculated results show that the $e_{31}$ or $d_{31}$  has small dependence on strain. For example,
at 4\%  strain, the $d_{31}$ of SbTeI (BiTeI)  becomes  -0.41 pm/V (-0.71 pm/V) from unstrained -0.37 pm/V (-0.66 pm/V), enhanced by 11\% (8\%).
A 2D material showing  piezoelectricity should be  a semiconductor,  except breaking inversion symmetry. Calculated results show that they all are semiconductors in considered strain range.
Thus, tensile strain can enhance  piezoelectric effects of Janus monolayer SbTeI and BiTeI.

In summary, the electronic structures, piezoelectric properties and carrier mobilities
 of the Janus  monolayer SbTeI and BiTeI are studied by using first-principles calculations.
The inversion asymmetry together with the SOC effect give rise to
Rashba-like splitting bands  with an indirect band gap of 0.89 eV for SbTeI and 0.67 eV for BiTeI.
The specific structural symmetry along with flexible mechanical properties induce  large in-plane
piezoelectric coefficient ($d_{11}$=12.95 pm/V for SbTeI and  $d_{11}$=8.20 pm/V for BiTeI) and an additional
out-of-plane piezoelectric coefficient ($d_{31}$=-0.37 pm/V for SbTeI and  $d_{31}$=0.66 pm/V for BiTeI).
Their carrier mobilities are attained with pronounced anisotropy, and the  hole mobilities are higher than electron ones. The hole carrier mobility of the BiTeI monolayer along the armchair direction  could be as high as 1319 $\mathrm{cm^2V^{-1}s^{-1}}$.
These findings can stimulate experimental
works to explore piezoelectric and electronic  applications of Janus  monolayer SbTeI and BiTeI.

\begin{acknowledgments}
This work is supported by the Natural Science Foundation of Shaanxi Provincial Department of Education (19JK0809). We are grateful to the Advanced Analysis and Computation Center of China University of Mining and Technology (CUMT) for the award of CPU hours and WIEN2k/VASP software to accomplish this work.
\end{acknowledgments}

\end{document}